\input{aipcheck}

\documentclass[
    ,final            
  ]
  {aipproc}

\layoutstyle{8x11single}

\begin{document}

\title{A preliminary analysis of $\eta^\prime\to\eta\pi\pi$ in chiral theories\footnote{
Talk given at HADRON 2009: XIII International Conference on Hadron Spectroscopy,
Florida, USA, 11/29-12/04/2009}}


\author{R.~Escribano}{
  address={Grup de F\'{\i}sica Te\`orica and IFAE, Universitat Aut\`onoma de Barcelona,\\
                    E-08193 Bellaterra (Barcelona), Spain}
}

\begin{abstract}
Preliminary results for the Dalitz plot distribution of $\eta^\prime\to\eta\pi\pi$ decays
in the frameworks of Large-$N_c$ Chiral Perturbation Theory and Resonance Chiral Theory
are given.
We hope our results to be of some relevance for the present and forthcoming analysis of these decays at GAMS, CLEO, VES, KLOE-2, Crystal Ball, Crystal Barrel, WASA, and BES-III.
\end{abstract}

\maketitle


\section{Introduction}

The decays $\eta^\prime\to\eta\pi\pi$ are interesting for several reasons.
First, due to the quantum numbers of the pseudoscalar mesons involved, if the decay proceeds through resonances these can be mostly of scalar nature.
$G$-parity prevents vectors to contribute.
Therefore, this decay is specially intended for the analysis of the properties of the $f_0(600)$
(or $\sigma$) resonance even though the $a_0(980)$ is also present and indeed dominant.
Second, the presence of $\eta$ and $\eta^\prime$ in this reaction is ideal for studying the mixing properties of both mesons.
Third, and more general, this decay allows to test Chiral Perturbation Theory (ChPT) and its possible extensions such as Large-$N_c$ ChPT and Resonance Chiral Theory (RChT).
For all that, precision measurements on $\eta$ and $\eta^\prime$ would be very helpful and provide useful information in our understanding of low energy QCD.
In particular, their decays are perfectly suited to study symmetries and symmetry breakings in QCD and the simultaneous treatment of both the $\eta$ and $\eta^\prime$ imposes tighter constraints on theoretical approaches than just the $\eta$.
Accordingly, there is at present an intense activity in studying these processes.
Recently, the GAMS-$4\pi$ and VES Collaborations have measured the related Dalitz plot parameters
(GAMS-$4\pi$ for the $\eta^\prime\to\eta\pi^0\pi^0$ channel \cite{Blik:2009zz} and
VES for the $\eta^\prime\to\eta\pi^+\pi^-$ one \cite{Dorofeev:2006fb})
complementing older results reported by an early GAMS Coll.~\cite{Alde:1986nw} and
CLEO \cite{Briere:1999bp}.
In the isospin limit the values of the Dalitz plot parameters should be the same,
however the experimental measurements show some discrepancies among them.
Then, there is a need of improvement and new measurements of theses parameters are foreseen at KLOE-2, Crystal Ball, Crystal Barrel and maybe WASA with improved statistics and a good understanding of the systematical errors.
At KLOE the $\eta^\prime$ is produced via the process $e^+e^-\to\Phi$ followed by
$\Phi\to\eta^\prime\gamma$.
For the $\eta^\prime\to\eta\pi^+\pi^-$ decay channel around 21K events are now on tape.
The background is seen to be very low and these data can easily be used to determine the Dalitz plot parameters.
The increased luminosity at KLOE-2 (a factor 3) will open new possibilities for $\eta^\prime$ studies.
For instance, a Monte Carlo simulation of the $\eta^\prime\to\eta\pi^+\pi^-$ process shows that the detector has a good sensitivity to $\sigma$ \cite{Venanzoni:2010ix}.
At Crystal Ball, the $\eta^\prime$ is photo produced, $\gamma p\to\eta^\prime p$,
and can be identified, e.g., from its decay $\eta^\prime\to\eta\pi^0\pi^0\to 6\gamma$.
About 10K events were expected during 2009 for the $\eta^\prime\to\eta\pi^0\pi^0$ process.
At WASA, large samples of $\eta^\prime$ will be produced in the reaction $pp\to pp\eta^\prime$
as soon as the new detector is developed.
Instead, the reaction $pd\to {}^3\mbox{He}\eta^\prime$ is not very suitable for $\eta^\prime$ decay studies.
Finally, about 27 million and 12 million decay events could be detected at BES-III each year for
$\eta^\prime\to\eta\pi^+\pi^-$ and $\eta^\prime\to\eta\pi^0\pi^0$, respectively.
On the theory side, the $\eta^\prime\to\eta\pi\pi$ decays have been studied within an effective chiral Lagrangian approach in which the lowest lying scalar mesons are combined into a possible
nonet \cite{Fariborz:1999gr} and, more recently, within the framework of $U(3)$ chiral effective field theory in combination with a relativistic coupled-channels approach \cite{Borasoy:2005du}.
Here, we present tentative results for the Dalitz plot parameters of the decays
$\eta^\prime\to\eta\pi\pi$ in the frameworks of Large-$N_c$ Chiral Perturbation Theory and Resonance Chiral Theory.
This contribution is based on a more complete analysis that will be submitted in the
near future \cite{wip} (see also Ref.~\cite{Masjuan:2009qy}).

\section{Dalitz plot parameterization}
\label{DP}
The Dalitz plot distribution for the charged decay channel is described by the following two variables:
\begin{equation}
\label{DPpar}
X=\frac{\sqrt{3}}{Q}(T_{\pi^+}-T_{\pi^-})\ ,\qquad
Y=\frac{m_\eta+2m_\pi}{m_\pi}\frac{T_\eta}{Q}-1\ ,
\end{equation}
where $T_{\pi,\eta}$ denote the kinetic energies of mesons in the $\eta^\prime$ rest frame and
$Q=T_\eta+T_{\pi^+}+T_{\pi^-}=m_{\eta^\prime}-m_\eta-2m_\pi$.
The squared absolute values of the two decay amplitudes (charged and neutral) are expanded around the center of the corresponding Dalitz plot in order to obtain the Dalitz slope
parameters\footnote{
The parameterization in Eq.~(\ref{A2}) had been proposed in Ref.~\cite{Montanet:1994xu}
with an extra term $eXY$.
The analysis of Ref.~\cite{Dorofeev:2006fb} included this term in their fits and found that parameter $e$ is consistent with zero.}
\cite{Dorofeev:2006fb}:
\begin{equation}
\label{A2}
|A(X,Y)|^2=|{\cal N}|^2[1+aY+bY^2+cX+dX^2]\ ,
\end{equation}
where $a, b, c$ and $d$ are real parameters and $|{\cal N}|^2$ is a normalization factor.
For the charged channel odd terms in $X$ are forbidden due to charge conjugation symmetry,
while for the neutral $c=0$ from symmetry of the wave function.
The Dalitz plot parameters may not be necessarily be the same for charged and neutral decay channels.
However, in the isospin limit they should be the same.
A second parameterization is the linear one \cite{Amsler:2008zzb}:
\begin{equation}
\label{A2linear}
|A(X,Y)|^2\propto |1+\alpha Y|^2+cX+dX^2\ ,
\end{equation}
where $\alpha$ is a complex parameter.
Comparison with the general fit gives $a=2\mbox{Re}(\alpha)$ and
$b=\mbox{Re}^2(\alpha)+\mbox{Im}^2(\alpha)$.
Both parameterizations are equivalent if $b>a^2/4$.

\section{Experimental data}
\label{expdata}
\begin{table}
\begin{tabular}{cccccc}
\hline
Parameter &
GAMS-4$\pi$ \cite{Blik:2009zz} & Theory \cite{Borasoy:2005du} &
VES \cite{Dorofeev:2006fb} & Theory \cite{Borasoy:2005du} &
This work \\
\hline
$a$ & $-0.066\pm 0.016\pm 0.003$ & $-0.127\pm 0.009$
& $-0.127\pm 0.016\pm 0.008$ & $-0.116\pm 0.011$ & -0.121\\
$b$ & $-0.063\pm 0.028\pm 0.004$ & $-0.049\pm 0.036$
& $-0.106\pm 0.028\pm 0.014$ & $-0.042\pm 0.034$ & -0.015\\
$c$ & $-0.107\pm 0.096\pm 0.003$ & ---
& $+0.015\pm 0.011\pm 0.014$ &  --- & ---\\
$d$ & $+0.018\pm 0.078\pm 0.006$ & $+0.011\pm 0.021$
& $-0.082\pm 0.017\pm 0.008$ & $+0.010\pm 0.019$ & -0.075\\
\hline \\
\end{tabular}
\caption{Dalitz slope parameters (experiment and theory) for
$\eta^\prime\to\eta\pi^0\pi^0$ (second and third columns) and
$\eta^\prime\to\eta\pi^+\pi^-$ (fourth and fifth columns), respectively.
The isospin-symmetric values of Eq.~(\ref{DPRChTsubleading}) are shown in the last column.}
\label{tablepar}
\end{table}

The latest available experimental information on the Dalitz slope parameters is summarized in
Table \ref{tablepar}.
The analysis by the GAMS-4$\pi$ Collaboration is based on approximately 15000 events
\cite{Blik:2009zz}.
Note that $b$ is negative here and the fit is thus not compatible with a linear fit.
If a fit is done with the linear parameterization one gets $\alpha=-0.042\pm 0.008$,
which is in agreement with an early measurement from GAMS based on 5400 events that gave
$\alpha=-0.058\pm 0.013$ (assuming $\mbox{Im}(\alpha)=0$ and $c=0$) \cite{Alde:1986nw}.
Both analyses are for the $\eta^\prime\to\eta\pi^0\pi^0$ channel.
Regarding the $\eta^\prime\to\eta\pi^+\pi^-$ decay, the result from CLEO based on 6700 events yields $\alpha=-0.021\pm 0.025$ (assuming $\mbox{Im}(\alpha)=0$, $c=0$ and $d=0$)
using the same linear fit as for the neutral decay channel \cite{Briere:1999bp}.
The VES analysis is based on roughly 13600 events obtained from the charge-exchange reaction
$\pi^-p\to\eta^\prime n$ and 6500 events from the diffractive-like production reaction
$\pi^-N\to\eta^\prime\pi^- N$ \cite{Dorofeev:2006fb}.
A fit using the combined data sets from VES gave the parameter values shown in
Table \ref{tablepar}.
Again, $b$ is found to be negative and thus incompatible with a linear fit,
while $c$ is consistent with zero.
The previous work supersedes a first study \cite{Amelin:2005rz},
where the $\eta\pi^+\pi^-$ Dalitz plot has been investigated with a sample of approximately
7000 events obtaining $\alpha=-0.072\pm 0.012\pm 0.006$
---this is the real part of $\alpha$ while $\mbox{Im}(\alpha)=0.0\pm 0.1\pm 0.0$---
in the linear and $a=-0.120\pm 0.027\pm 0.015$ in the nonlinear parametrization.
The $C$-violation parameter was compatible with zero, $c=0.021\pm 0.024$.
In average, $\alpha=-0.059\pm 0.011$ \cite{Amsler:2008zzb}.
Finally, a measurement of $\alpha$ not included in the average is reported in
Ref.~\cite{Kalbfleisch:1974ku}, where $\alpha=-0.08\pm 0.03$
(assuming $\mbox{Im}(\alpha)=0$ and $c=0$) with about 1400 events.
For the general parameterization, we see from Table \ref{tablepar} that there is some tension in all  the parameters.
Notice, however, that the theory model results of Ref.~\cite{Borasoy:2005du}
for the charged and neutral decay channels are compatible among themselves.
While there is agreement in $a$ between the VES fitted value and the result of
Ref.~\cite{Borasoy:2005du} there is not such when compared to the GAMS-$4\pi$ value.
The same happens to $d$ but this time the GAMS-$4\pi$ and the theory model results agree,
although are in conflict with the VES reported value.
The case of $b$ is less severe.
The VES value is compatible with the others only at 2$\sigma$.
Finally, there is a variation between the measured values of $c$.
Nevertheless, the large statistical error of the GAMS-$4\pi$ result makes this statement not conclusive.

\section{Chiral theories}
\label{theories}
In this contribution we study the decays $\eta^\prime\to\eta\pi\pi$ in two different chiral theories:
Large-$N_c$ Chiral Perturbation Theory (Large-$N_c$ ChPT) and
Resonance Chiral Theory (RChT).
Both theories emerge from Chiral Perturbation Theory with the following modifications.
ChPT is based on the spontaneous symmetry breaking of the $SU(3)_L\times SU(3)_R$
chiral symmetry exhibited by QCD at low energies down to $SU(3)_V$.
As a result, an octet of very light pseudoscalar bosons $(\pi, K,\eta_8)$ with small masses compared to a typical hadronic scale such as $M_\rho$ appear.
The pseudoscalar singlet $\eta_0$ is not included in ChPT as an explicit degree of freedom since the $U(1)$ axial anomaly prevents this state from being considered a ninth pseudo-Goldstone boson.
However, in the large-$N_c$ limit the $U(1)_A$ anomaly is absent and the original chiral symmetry of QCD in the massless limit is enlarged to $U(3)_L\times U(3)_R$.
Large-$N_c$ ChPT is organized as a systematic expansion in powers of momenta, quark masses and $1/N_c$
---$p^2\sim m_q\sim 1/N_c={\cal O}(\delta)$---
describing the low energy interactions of $\pi, K, \eta$ and $\eta^\prime$
\cite{Kaiser:2000gs}.
Then, in principle, it is perfectly suited for the study of $\eta$ and $\eta^\prime$ hadronic decays.
In the previous two frameworks lowest-mass resonances are not yet incorporated
although it seems rather natural to expect an important impact of these resonances on the physics of the lightest pseudoscalar bosons.
A systematic analysis of the role of resonances in the ChPT Lagrangian, the so-called RChT,
was performed in Ref.~\cite{Ecker:1988te}.
In the following two sections we give tentative results for the Dalitz slope parameters predicted by these two theories.

\subsection{Large-$N_c$ Chiral Perturbation Theory}
\label{largeNcChPT}
At lowest order (lo) in Large-$N_c$ ChPT the amplitude for $\eta^\prime\to\eta\pi^+\pi^-$
is written as
\begin{equation}
\label{lo}
A_{\rm lo}=\frac{m_\pi^2}{2F^2}\sin(2\phi)\ ,
\end{equation}
where $F=F_\pi$ at this order and
$\phi$ is the $\eta$-$\eta^\prime$ mixing angle in the quark-flavor basis.
Taking $F_\pi=92.3$ MeV and the latest value $\phi=40.4^\circ$ from KLOE
\cite{Ambrosino:2009sc}
one gets
$B(\eta^\prime\to\eta\pi^+\pi^-)\sim 10^{-2}$
which is very small compared to the experimental value
$B(\eta^\prime\to\eta\pi^+\pi^-)=(44.6\pm 1.4)\times 10^{-2}$ \cite{Amsler:2008zzb}.
The reason for such a small value is the appearance of $m_\pi^2$ in Eq.~(\ref{lo}),
\textit{i.e.}~a chirally suppressed contribution.
Moreover, the Dalitz plot distribution is constant in clear disagreement with the measurement of unambiguous slope parameters.
At next-to-leading order (nlo) the $\eta^\prime\to\eta\pi^+\pi^-$ amplitude involves the low-energy constants $L_5, L_8$ and the combination $3L_2+L_3$ together with the OZI-rule violating parameter $\Lambda_2$ \cite{Kaiser:2000gs}.
The contributions from $L_5, L_8$ and $\Lambda_2$ are all proportional to $m_\pi^2$
and therefore chirally suppressed.
The contribution from $3L_2+L_3$ is seen to be dominant and indeed, at this order, contains all the dynamical information encoded in the Dalitz parameters.
For the numerical analysis, we use values for the pseudoscalar decay constants and mixing angles from Ref.~\cite{Escribano:2005qq}
and the low-energy constants from Refs.~\cite{Bijnens:1994qh,Ecker:1994gg}.
The values for the mixing parameters, which are extracted from phenomenology,
are in reasonable agreement with the predictions of Large-$N_c$ ChPT \cite{Leutwyler:1997yr}.
After expanding the exact expression of $|A_{\rm nlo}|^2$ in terms of $X$ and $Y$
one finds the following relevant fact:
$|A_{\rm nlo}|^2$ only depends on $X^2$ in the $m_\pi\to 0$ limit
(keeping $m_{\eta,\eta^\prime}\neq 0$).
This implies that terms proportional to $Y$ carry extra powers of $m_\pi$,
thus making these terms numerically comparable to those of order $X^2$.
As a consequence, to be consistent our Dalitz expansion needs to include terms of order
$X^2 Y$ and $X^4$ denoted by $\kappa_{21}$ and $\kappa_{40}$, respectively.
We also find combinations of slope parameters, such as $a/d$ and $\kappa_{40}/\kappa_{21}$,
which are independent of the low-energy constants and given as functions of the pseudoscalar masses alone.
Preliminary values for the Dalitz parameters obtained from this Large-$N_c$ expansion are
\begin{equation}
\label{DPLNcChPT}
a[Y]=-0.282\ ,\quad
d[X^2]=-0.083\ ,\quad
b[Y^2]\sim 10^{-3}\ ,\quad
\kappa_{21}[X^2Y]=0.012\ ,\quad
\kappa_{40}[X^4]\sim 10^{-3}\ .
\end{equation}
As seen, the value for $d$ is in accord with the VES result while the values for $a$ and $b$ are not.
This latter may be interpreted as an indication of the important role played by the $\sigma$ meson in this process whose contribution is encoded in the $Y$ terms.
Since the $\sigma$ pole is close to the allowed kinematical region
a truncated expansion in $Y$ would not be able to reproduce their effects.
On the contrary, the parameter $d$ encodes the contribution of the $a_0$ which is kinematically suppressed and thus well represented by the expansion in $X$.
For those reasons, a better description of the role played by scalar resonances in
$\eta^\prime\to\eta\pi\pi$ decays is desirable and this is the subject of RChT.

\subsection{Resonance Chiral Theory}
\label{RChT}
In this chiral theory the $\eta^\prime\to\eta\pi\pi$ decays are driven by the exchange of the
$\sigma$ and $f_0$ in the $s$-channel and the $a_0$ in the $t$- and $u$-channel.
The contribution of scalar mesons in the RChT Lagrangian
with a minimal number of resonance fields is represented by two operators whose
coupling constants are $c_d$ and $c_m$ \cite{Ecker:1988te}.
The largest contribution to the $\eta^\prime\to\eta\pi\pi$ amplitude comes from the $c_d$ terms,
which are proportional to the external momenta, while the rest is proportional to $m_\pi^2$ and then suppressed.
At low energies, when $s,t,u\ll M_S^2$,
the amplitude in the RChT framework recovers that of Large-$N_c$ ChPT up to subleading contributions in $1/N_c$.
After using the low-energy constants relations
$3L_2+L_3=c_d^2/2M_S^2$, $L_5=c_d c_m/M_S^2$ and $L_8=c_m^2/2M_S^2$,
and the large-$N_c$ result $\Lambda_2=0$,
the two amplitudes match.
For the numerical analysis, we use the same set of input values as for the Large-$N_c$ ChPT case
together with $M_S=980$ MeV for the scalar multiplet mass.
We also use $c_d=32\pm 13$ MeV and $c_m=42\pm 7$ MeV from Ref.~\cite{Ecker:1988te}.
They are obtained from the phenomenological values
$L_5^r(M_\rho)=(1.4\pm 0.5)\cdot 10^{-3}$ and $L_8^r(M_\rho)=(0.9\pm 0.3)\cdot 10^{-3}$ assuming scalar resonance saturation.
These values of $c_d$ and $c_m$ allow for a prediction of $B(\eta^\prime\to\eta\pi^+\pi^-)$
in agreement with experiment.
However, since the total decay width is largely proportional to $c_d^4$
the propagated error is huge.
Therefore, we prefer to fix $c_d$ from the experimental branching ratio \cite{Amsler:2008zzb},
thus obtaining $c_d=28.9\pm 0.2$ MeV,
and $c_m$ from the short-distance constraint in the single-resonance approximation
$4c_d c_m=f^2\approx f_\pi^2$ \cite{Jamin:2001zq}.
The values for the Dalitz parameters one gets in this case are
\begin{equation}
\label{DPRChT}
a[Y]=-0.116\ ,\quad
d[X^2]=-0.054\ ,\quad
b[Y^2]\sim 10^{-3}\ ,\quad
\kappa_{21}[X^2Y]=-0.006\ ,\quad
\kappa_{40}[X^4]\sim 10^{-3}\ .
\end{equation}
Now, the values of $a$ and $d$ agree with the VES results while $b$ is still very small.
This behavior may indicate that $\sigma$ meson effects are already accounted for
but not yet in precise form.
So far, subleading effects in large-$N_c$ such as final state interactions
(making the $\sigma$ a broad resonance) or the use of different scalar masses within the multiplet are not considered.
Different ways of including these effects haven been studied in Ref.~\cite{wip}.
For instance, the inclusion of a $\sigma$-$f_0$ mass difference
along with a self-energy in the $\sigma$ propagator that reproduces the correct $\sigma$ pole
in the complex plane (the same is done for $a_0$ and $f_0$) gives
\begin{equation}
\label{DPRChTsubleading}
a[Y]=-0.121\ ,\quad
d[X^2]=-0.075\ ,\quad
b[Y^2]=-0.015\ ,\quad
\kappa_{21}[X^2Y]=-0.008\ ,\quad
\kappa_{40}[X^4]\sim 10^{-3}\ .
\end{equation}
We see that $a$ and $d$ are in nice agreement with the VES results,
whereas the prediction for $b$, although increased in one order of magnitude,
is still far from its experimental value.
For completeness, we display in Table \ref{tablepar}
our former results for the parameters $a$, $b$ and $d$.

\section{Summary and discussion}
In this contribution we have computed preliminary values for the Dalitz slope parameters related to $\eta^\prime\to\eta\pi\pi$ decays in the frameworks of
Large-$N_c$ Chiral Perturbation Theory, at lowest and next-to-leading orders, and
Resonance Chiral Theory in the leading $1/N_c$ approximation.
We have seen that next-to-leading order Large-$N_c$ ChPT fails to reproduce the Dalitz parameters associated with scalar exchanges in the $s$-channel,
thus indicating the relevance of $\sigma$ meson effects in this channel.
The contributions from the $t$- and $u$-channel are, however, well reproduced.
The prediction for the Dalitz distribution is improved in the context of RChT where scalar resonance effects are included explicitly and not integrated out in the low-energy constants.
We have then seen that the contribution of the $a_0$ is dominant but that of the $\sigma$ is essential to reach agreement at the current precision level.
The situation gets even better as soon as subleading effects in $1/N_c$ are considered.
The incorporation of a $\sigma$-$f_0$ mass splitting and some final state interactions through 
the modification of the $\sigma$ propagator seem to be unavoidable in order to match
the $a$ and $d$ parameters of the Dalitz plot expansion.
The $b$ parameter is, however, not yet reproduced.
For the time being, we do not foresee how to make this parameter to agree.
This is in contrast with the predictions of the $U(3)$ chiral effective field theory with coupled channels where $a$ and $b$ are nicely reproduced but $d$ is not.
Likewise, current experimental results show some tension among themselves,
especially in the case of the $a$ parameter.
We hope that forthcoming analysis of the $\eta^\prime\to\eta\pi\pi$ decays at different facilities
will clarify the present situation and allow for a distinction between different chiral theory predictions.


\begin{theacknowledgments}
I would like to express my gratitude to the HADRON 2009 Organizing 
Committee for the opportunity of presenting this contribution, and for the 
pleasant and interesting workshop we have enjoyed.
This work was supported in part by
the Ministerio de Ciencia e Innovaci\'on under grant CICYT-FEDER-FPA2008-01430, 
the EU Contract No.~MRTN-CT-2006-035482, ``FLAVIAnet'',
the European Commission under the 7th Framework Programme through the 'Research Infrastructures' action of the 'Capacities' Programme. Call: FP7-INFRASTRUCTURES-2008-1, Grant Agreement N. 227431,
the Spanish Consolider-Ingenio 2010 Programme CPAN (CSD2007-00042), and
the Generalitat de Catalunya under grant SGR2009-00894.
\end{theacknowledgments}

\end{document}